\documentclass[letterpaper]{article}
\usepackage{aaai}
\usepackage[dvips]{graphicx}

\usepackage{times}
\usepackage{helvet}
\usepackage{courier}
\usepackage[latin1]{inputenc}
\usepackage{rotating}

\nocopyright

\title{Herschel Mission Planning Software}
\author{
Jon Brumfitt, Pedro G\'omez-Alvarez, Antonio Villacorta, Pedro Garc\'{\i}a-Lario, Rosario Lorente, Laurence O'Rourke\\
Herschel Science Centre, European Space Astronomy Centre (ESAC),\\
European Space Agency (ESA), P.O. Box 78, Villanueva de la Cañada, \\
28691, Madrid, Spain}
\begin{document}
\maketitle
\begin{abstract}
The mission planning software developed for the Herschel Project is presented: The Herschel Inspector and mid/Long-term Scheduler (HILTS) and the short-term scientific Mission Planning system (SMPS).
\end{abstract}

\section{Introduction}
Herschel is an ESA cornerstone observatory mission launched on 14 May 2009. 
Herschel covers the wavelength range from 55 to 672 microns with three instruments: HIFI, PACS and SPIRE (\citeauthor{pilbratt2010} 
\citeyear{pilbratt2010} and references therein).
Planning, visualization and inspection capabilities are important in any observatory. Cryogenic space 
observatories such as Herschel, which has an estimated lifetime of 3.5-4 years, call for additional efforts to 
maximize the observatory scientific return.

This paper starts with a brief description of those Herschel mission aspects relevant to mission planning, including proposals, instruments and constraints. It will then describe the two tools that comprise the Herschel Mission planning software:  the Herschel Inspector and Long-Term Scheduler (HILTS) and the Scientific Mission Planning System (SMPS).

HILTS was initially conceived to assist Herschel medium and long-term planning. The tool is also useful 
to assess the mission's past, present and future status. Short-term mission planning for a given 
Operational Day\footnote{The so-called operational day (OD) figure is the number of days elapsed from beginning of the
Herschel mission} (OD) is executed using the Herschel Scientific Mission Planning System (SMPS) \cite{brumfitt2005}, 
which generates satellite telecommands that are uplinked to Herschel on a daily basis. 
Both HILTS and SMPS have been developed using a common Java object-oriented framework which implements basic astrometric, graphical,
timing, pointing, etc functionality. This framework was initially developed for SMPS and then reused by HILTS.

\section{The Mission}
\begin{figure}[!ht]
  	\centering
  	\includegraphics[scale=0.15]{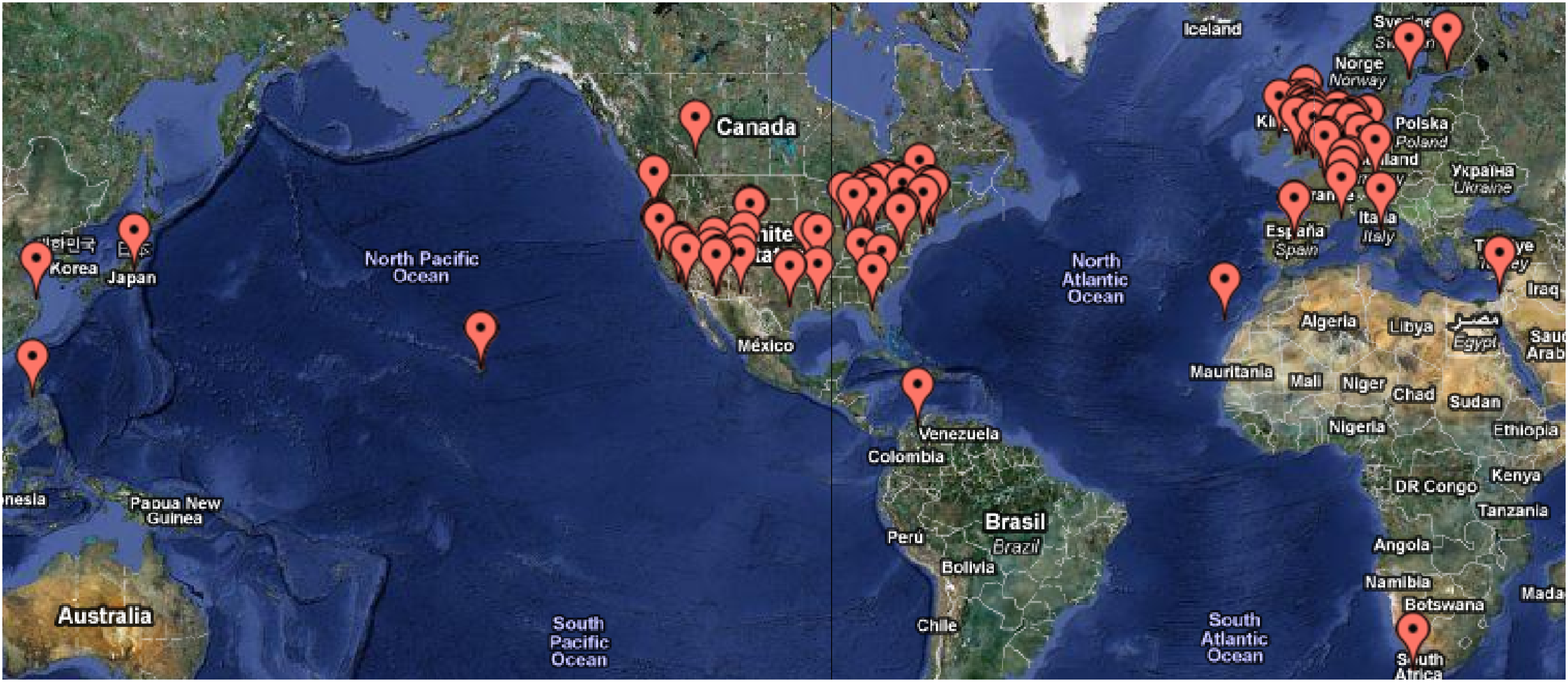}
  	\caption{Geographical distribution of Herschel's proposals}
	\label{fig:GeographicalDistribution}
\end{figure}

Herschel's operational database is populated by approximately 30,000 observation requests pertaining to 411 science proposals, 
from around 300 PIs (see figure \ref{fig:GeographicalDistribution} for their geographic distribution). 
There are several factors impacting Herschel scheduling: helium optimization, slews minimization, 
proposal completion, scientific grades, Targets of Opportunity (ToOs) and operational issues. Herschel 
has also thermal and communication constraints: the observatory attitude is constrained by the 
(anti)Sun, Earth, Moon and some planets; the observatory needs also to communicate with the ground station 
every 24 hours, during the so-called Daily Telecommunication Period (DTCP).

\subsection{Instruments}

The Herschel spacecraft has three instruments, HIFI, PACS and SPIRE. However, for mission planning purposes, 
each instrument is treated as a number of sub-instruments, as follows:
\begin{itemize}
	\item HIFI bands [1A,1B,...,7A,7B]
	\item PACS photometry
	\item PACS spectrometry
	\item SPIRE photometry
	\item SPIRE spectrometry
	\item SPIRE/PACS parallel mode
\end{itemize}

Each OD is typically assigned to a single sub-instrument in order to improve efficiency. This reduces the overheads in 
switching between sub-instruments.  In the case of HIFI, several bands are often scheduled in an OD.  The largest overhead is recycling 
the liquid Helium cooler. The PACS and SPIRE instruments each have their own cooler system, which requires recycling separately, 
each taking approximately 3 hours. Once recycled, the instrument can be used for about two days. For PACS, recycling is only needed for photometry 
and not for spectrometry.  For example, having spent three hours recycling the PACS cooler, it is most efficient to perform only PACS 
photometry observations for the following two days. Similarly, after recycling the SPIRE cooler, the following two days should be devoted 
to SPIRE. In parallel mode, the PACS and SPIRE coolers can be recycled in parallel, taking about three hours, after which parallel mode 
observations should be performed for two days.

Since the spacecraft attitude is severely constrained during the three-hour DTCP, this time is less available to be used for scientific observations. 
Consequently, it is a good time to recycle the cooler and perform any other instrument preparation.

The other instrument preparation times are less significant than cooler recycling, but nevertheless make it desirable to keep to one 
sub-instrument per day. For example, preparing PACS for either photometry or spectrometry takes about 30 minutes (each), in addition 
to the cooler recycling time needed for photometry.

Instrument preparation operations, such as cooler recycling and detector curing, are known as engineering observations, as they do not 
perform a scientific measurement. They are scheduled like scientific observations but do not require any particular spacecraft attitude.

\subsection{Constraints}

\begin{figure}[!ht]
  	\centering
  	\includegraphics[scale=0.50]{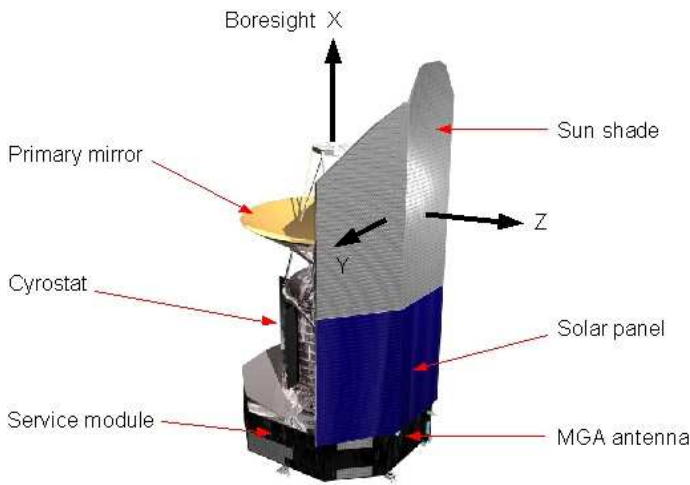}
  	\caption{Herschel spacecraft showing axis directions}
	\label{fig:mps_spacecraft}
\end{figure}

\begin{figure}[!ht]
  	\centering
  	\includegraphics[scale=0.45]{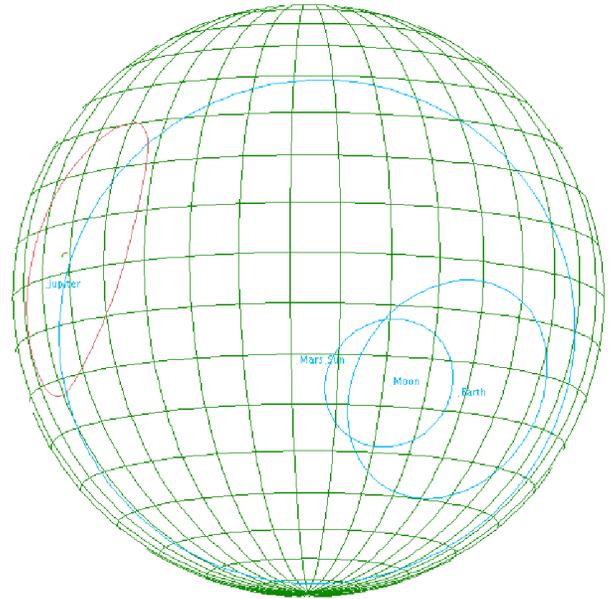}
  	\caption{Herschel pointing constraints: See the big solar and the two smaller earth and moon constraints as well as the elliptical MGA constraint, which is in effect 
	during the daily communication period (DTCP)}
	\label{fig:mps_constraints}
\end{figure}

The Herschel spacecraft is shown in figure \ref{fig:mps_spacecraft}. The solar panel is nominally positioned to face the Sun so as to generate 
sufficient power. In this configuration, the Sun shade and solar panel shield the telescope and cryostat from the Sun. The spacecraft may be 
maneuvered by $\pm 30$ degrees about the Y axis but by only $\pm 3$ degree about the X axis, in order to satisfy power and thermal considerations. 
The telescope boresight points along the X axis and therefore has a view of a circular band of sky, 60 degrees wide, covering about 50\% of the sky. 
The instruments impose small additional circular restrictions on the boresight around Mars, Jupiter and Saturn. Earth and Moon 
constraints normally lie within the Sun constraint. These constraints are illustrated in figure \ref{fig:mps_constraints}.

Communication with the ground station takes place in a 3-hour Daily Telecommunications Period (DTCP), during which the spacecraft is maneuvered to 
point its Medium Gain Antenna (MGA) at the Earth. This period is used to uplink telecommands and downlink the resulting telemetry from the previous 
day. In principle, scientific observations can be performed during DTCP, but only a restricted field of view.

The antenna constraint on the Z axis, during DTCP, results in an additional constraint on the boresight direction, when it is combined with the Sun 
constraint. When the Sun-spacecraft-Earth angle is greater than about 15 degrees, this takes the form of two roughly elliptical regions on the 
sky, one of which can be seen in figure \ref{fig:mps_constraints}.

\section{Herschel Inspector and mid/Long term Scheduler (HILTS)}
Although initially intended as a Long-Term scheduler, it was soon realized that it
 could be also useful for other important purposes. Amongst the features currently present in the tool are:
 \begin{itemize}
 	\item Automatic and manual medium/long-term scheduler
	\item Mission inspector: It gives a means to assess various aspects of the mission at any given time: visibility of sidereal and 
	solar system objects (at a given time, within the antenna constraint, etc), status of observations, cone searches, pointing history, 
	observation footprint, regions of the sky affected by stray light, etc.
	\item Query browser: it gives the means to perform arbitrary complex queries over more than 20 criteria.
	\item Statistical engine: it generates various types of statistics (see statistics section). 
	\item Duplications: With each observing call it is required to identify potential duplications between new observations and the existing ones, 
	helping to remove redundant science and thus maximizing scientific return.
	\item Catalog interface with IRAS, AKARI and user catalogs.
	\item VO-aware, to interoperate with other VO tools\footnote{Virtual Observatory http://www.ivoa.net/}.
 \end{itemize}

HILTS is a Java tool, whose main screen is divided into a set of panels (see figure \ref{fig:hilts}):
\begin{sidewaysfigure*}
  	\centering
  	\includegraphics[scale=0.35]{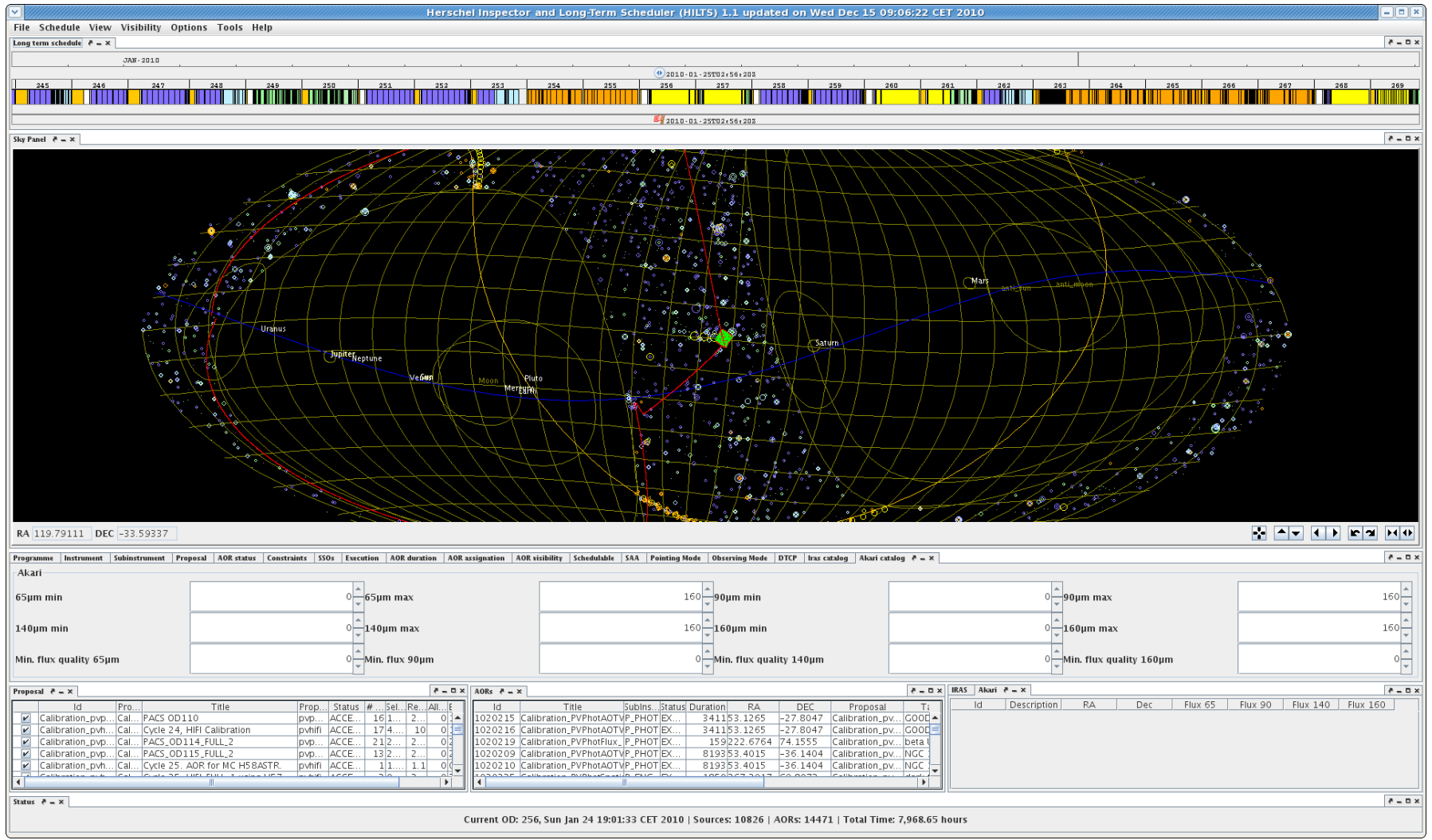}
  	\caption{HILTS main screen. From top to bottom and from left to right: The time panel with its sub-panels, the sky panel, the query panel
	 in which each tabbed panel represent a different query criterion, the proposal panel, the AOR panel, the catalog panel and 
	 the status panel}
	\label{fig:hilts}
\end{sidewaysfigure*}

\begin{itemize}
	\item \emph{Time panel:} It is in turn composed of a set of horizontal sub-panels: A simple Gregorian calendar; 
	a time selector, where the current time is selected; the OD sub-panel, where the OD divisions and select observation visibility 
	are represented; assigned observations sub-panel, where the scheduled and already executed observations are represented; 
	current observing block restrictions (groups of ODs preallocated to a given instrument mode) 
	and the available scheduling interval.
	\item \emph{Sky panel:} Visible observations and current constraints are represented in this panel. 
	The satellite pointing history for the current OD can also be plotted.
	\item \emph{Query panel:} Composed of multiple tabs, allowing arbitrary complex selections using 
	the available criteria: Observation programmes, instruments and instrument modes, request status, 
	Solar System Objects (SSOs), duration, etc.
	\item \emph{Proposal panel:} Current selected proposals are listed and can also be (de)selected.
	\item \emph{Requests panel:} Where the current selected observations are listed 
	\item \emph{Catalogs panel:} By default IRAS and AKARI catalogs. User catalogs can also be loaded. 
	\item \emph{Status panel:} General status information
\end{itemize}
All panels are interconnected with each other by means of the adoption of the Model-View-Controller (MVC) software pattern. 
For instance, when a new time is selected in the time panel, 
constraints and visible observations are updated simultaneously in the sky panel, while visible proposals and 
catalog objects are also updated in their respective panels. Selecting objects visible at a given time, 
is the default visibility selection. Other alternatives are available: (in)visibility during a time interval, 
during the DTCP, always visible, etc.

\subsection{Long-Term Scheduling}
\begin{figure*}
  	\centering
  	\includegraphics[scale=0.40]{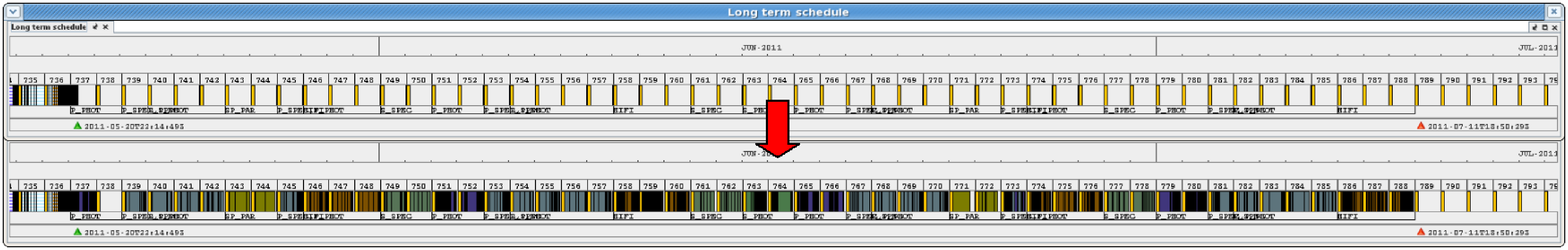}
  	\caption{Before and after an automatic scheduling run}
	\label{fig:scheduling}
\end{figure*}

HILTS supports both manual and automatic scheduling. The former, by simple drag-and-drop from the observation 
panel to the scheduled observations sub-panel. The tool automatically places the observation at the earliest 
time within the dropped OD, taking into account observing blocks, observations duration, configuration and slews, 
amongst other factors. The latter (see figure \ref{fig:scheduling}) is attained by first selecting a suitable 
interval of typically several months and one of the set of pluggable strategies. For instance, if the \emph{remaining visibility} 
strategy is selected, the tool will assign each of the visible observations by order of remaining visibility.\\
HILTS scheduling is typically an iterative process: starting with one of the available ``filler'' strategies and finishing 
with an optimization phase (a simulated annealing optimization is being developed). Once a satisfactory schedule is obtained,
 it can be exported to XML and loaded into the SMPS.

There are several factors impacting the quality of a schedule, amongst them: Helium and slew optimization, proposal and
programme completion, scientific priorities, sky homogenization, operational issues, ToOs, etc. All these factors generally compete with each other,
making it very difficult to generate the "perfect" schedule (provided such a thing can be defined) in one go. Instead, HILTS
 allows the effect of a given strategy to be simulated, giving significant information and helping the human being in making decisions.

\subsection{Statistics}
\begin{figure}[!ht]
  	\centering
  	\includegraphics[scale=0.30]{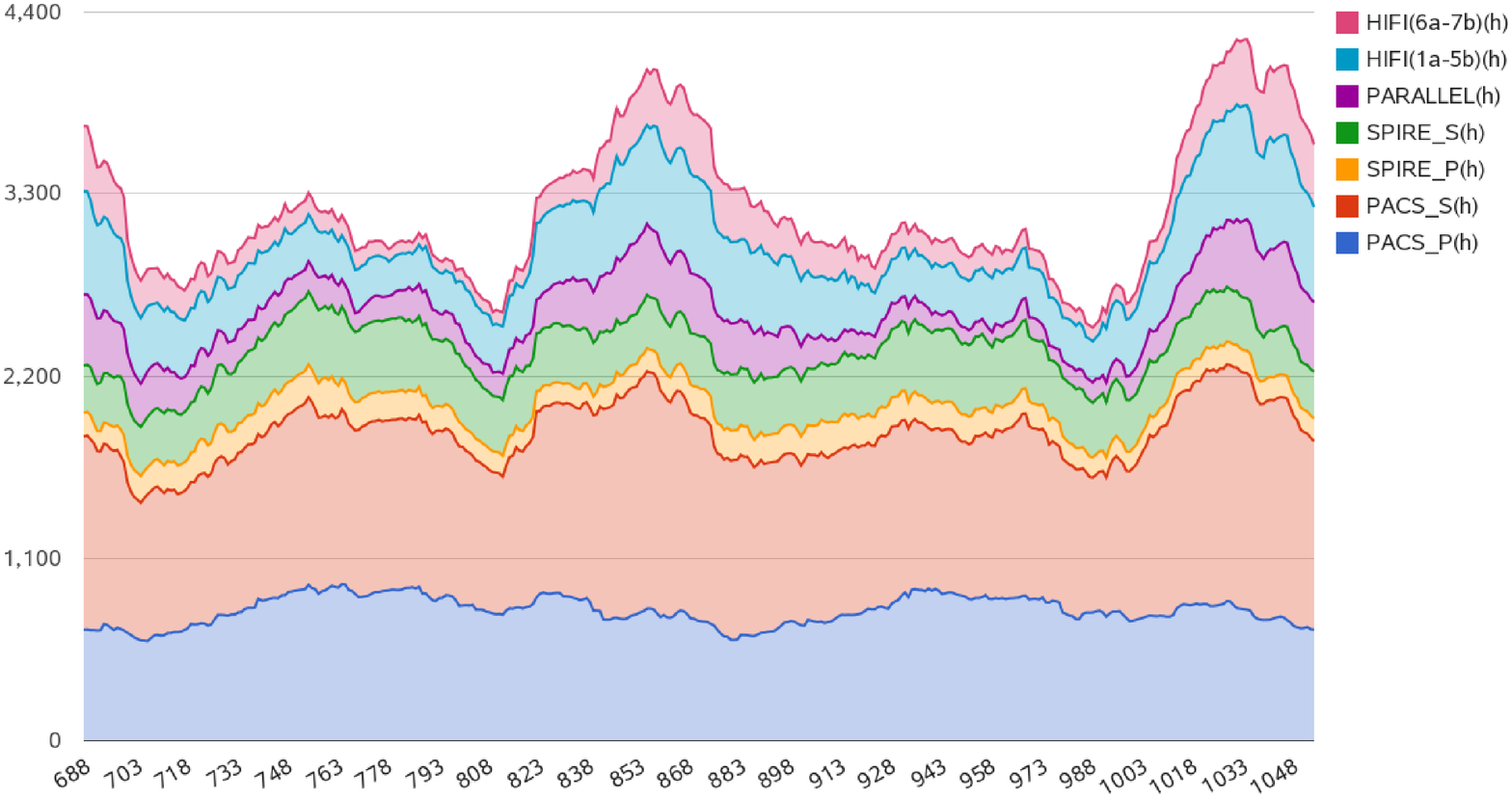}
  	\caption{Available time (hours) per day and sub-instrument throughout a year. The X axis represents the operational day}
	\label{fig:PressureReportAbsoluteStacked}
\end{figure}

\begin{figure}[!ht]
  	\centering
  	\includegraphics[scale=0.30]{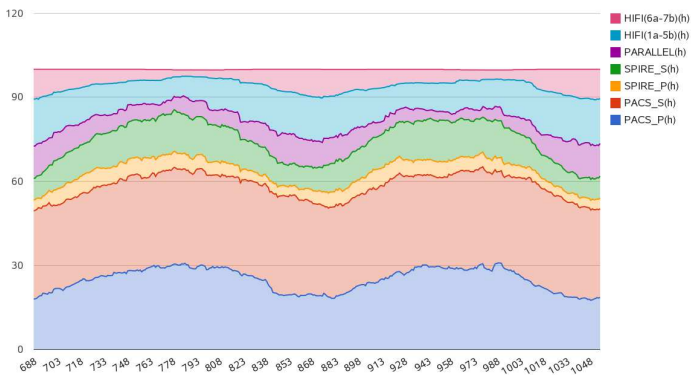}
  	\caption{Fraction of available time per sub-instrument and OD for a year}
	\label{fig:PressureReportRelativeStacked}
\end{figure}

\begin{figure}[!ht]
  	\centering
  	\includegraphics[scale=0.30]{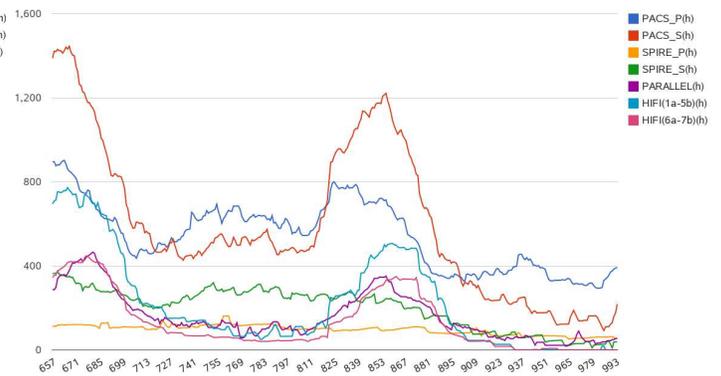}
  	\caption{Available time as a long-term schedule is generated.}
	\label{fig:ScheduleReportAvailableTime}
\end{figure}

\begin{figure}[!ht]
  	\centering
  	\includegraphics[scale=0.30]{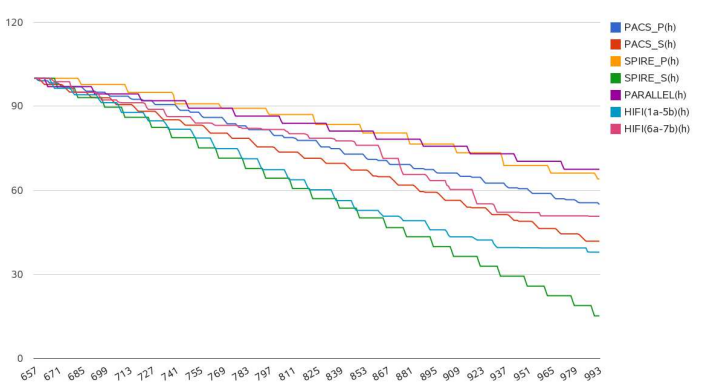}
  	\caption{Unobserved time (fraction of the initial) as a long-term schedule is generated.}
	\label{fig:ScheduleReportUnobservedPercentage}
\end{figure}

\begin{figure}[!ht]
  	\centering
  	\includegraphics[scale=0.25]{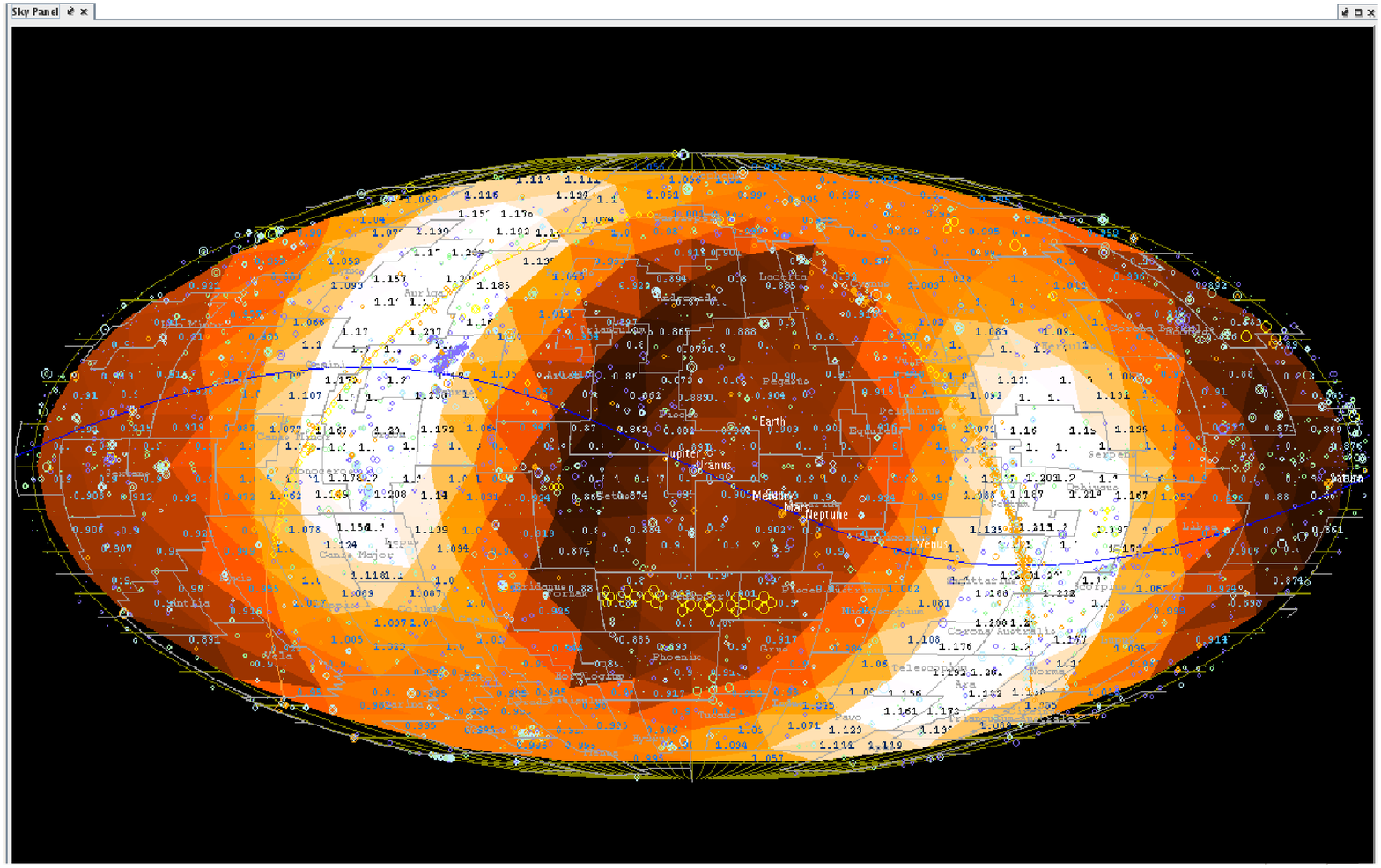}
  	\caption{All-Sky competition plot using a HTM tessellation}
	\label{fig:CompetitionTotal}
\end{figure}

\begin{figure}[!ht]
  	\centering
  	\includegraphics[scale=0.30]{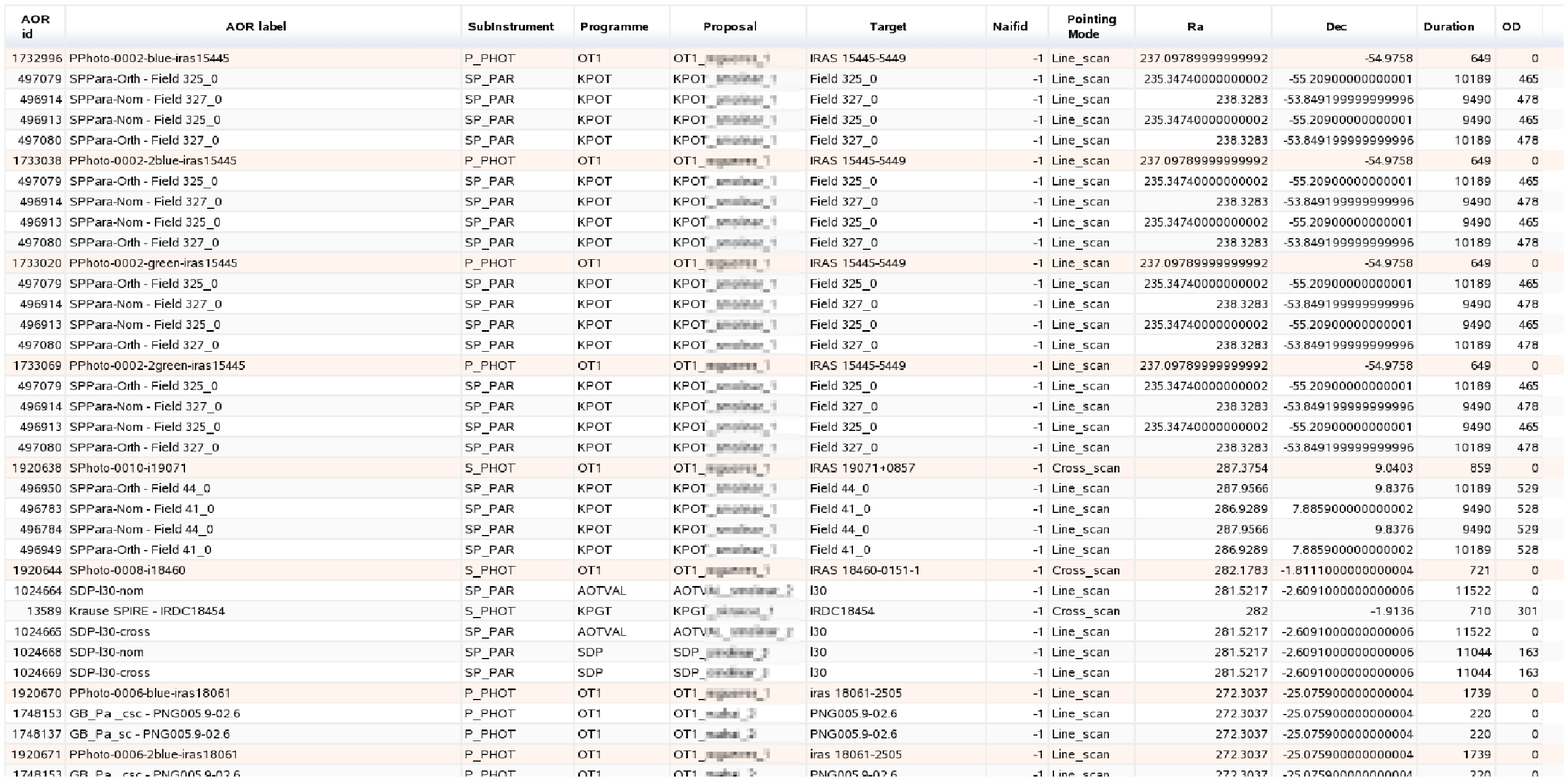}
  	\caption{Duplication analysis: Each group represents a set of potentially colliding observations}
	\label{fig:DuplicationsTable}
\end{figure}

\begin{figure}[!ht]
  	\centering
  	\includegraphics[scale=0.28]{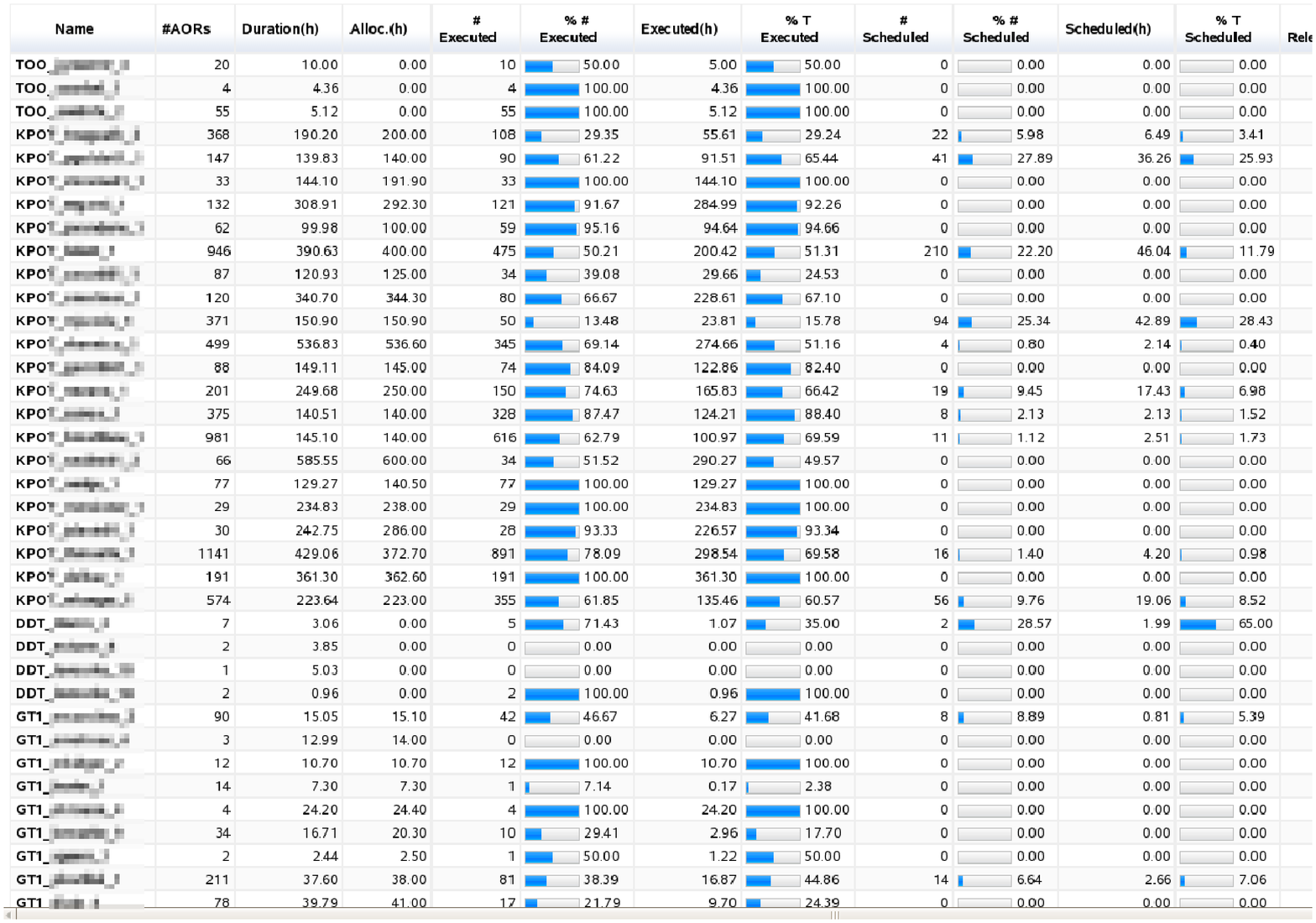}
  	\caption{Proposal reports}
	\label{fig:ProposalStatisticalReports}
\end{figure}

\begin{figure}[!ht]
  	\centering
  	\includegraphics[scale=0.12]{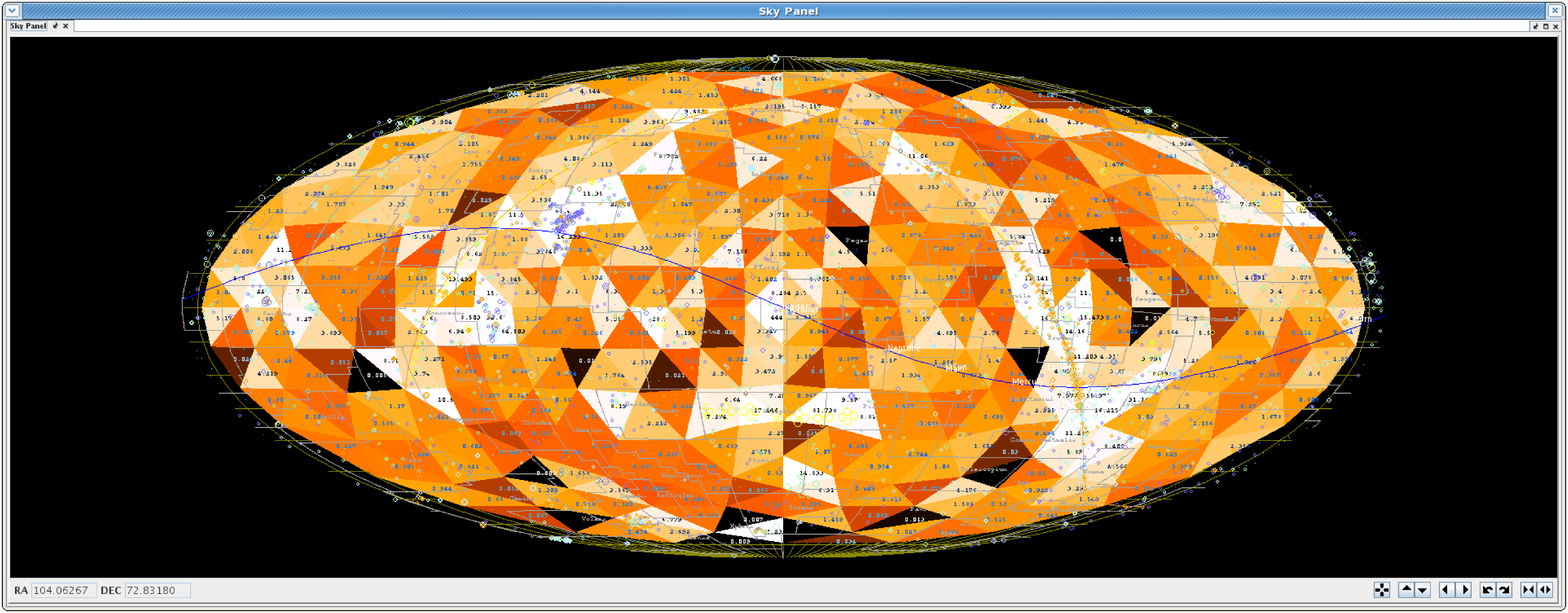}
  	\caption{Observing time density using HTM tessellation}
	\label{fig:TotalTimeDensity}
\end{figure}

HILTS can generate detailed statistics focused on several mission aspects, which help 
evaluate current mission status and thus retrofit observation strategies. Among the reports HILTS is capable of generating are: 
\begin{itemize}
	\item \emph{Mission status reports:} The completion status of each programme, proposal, instrument and instrument mode
	is generated on a weekly basis. This information is useful for the science and mission planning teams to assess the 
	overall mission status (see figure \ref{fig:ProposalStatisticalReports})
	
	\item \emph{Duplication reports:} With each open call of proposals it is necessary to determine which of the new observations could be potentially
	duplicating other observations in already accepted proposals. This is performed by the tool using two criteria: spatial and spectral overlapping 
	(see figure \ref{fig:DuplicationsTable}).
	
	\item \emph{Pressure reports:} HILTS can estimate the amount of observing time that is available during each OD in a given period of time 
	(see figure \ref{fig:PressureReportAbsoluteStacked} and \ref{fig:PressureReportRelativeStacked}). This information is useful to adequate the observing
	block schema to the available time for a given period of time.
	
	\item \emph{Scheduling reports:} When HILTS generates a long-term schedule it simultaneously generates statistical information of certain parameters such as:
	 the efficiency of each scheduled OD, the available time to schedule each OD, taking into account the already
	scheduled observations (see figure \ref{fig:ScheduleReportAvailableTime} and \ref{fig:ScheduleReportUnobservedPercentage})
	
	\item \emph{Density reports:} In order to represent spatial densities of any magnitude, the tessellation of celestial sphere using  
	the Hierarchical Triangular Mesh (HTM) has been used \cite{szalay2007}. See figure \ref{fig:TotalTimeDensity} for the global distribution of unobserved 
	time at the time of writing. Other plots (per instrument, programme, etc) are of course possible.
	
	\item \emph{Competition reports:} Represents the amount of observing time each region of the sky has to "compete" with, when it is visible (see
	 figure \ref{fig:CompetitionTotal}). It is interesting to note that this is in general decoupled with the aforementioned time densities: a
	 relatively shallowly-observed region can have a large competition figure, if has to share its visibility with densely-observed regions.
	
\end{itemize}

The AJAX Google presentation API\footnote{http://code.google.com/apis/visualization/documentation/gallery.html} has been extensively used to 
implement this functionality.

\subsection{Catalogs and Virtual Observatory}
HILTS is also able to interact with on-line catalogs from Vizier \cite{vizier2006}. Specially relevant catalogs such as IRAS and AKARI 
can be also filtered by flux in the query panel. A synthetic catalog of IR sources for the selection of candidate ``filler observations'' is also 
included.
\begin{figure}[!ht]
  	\centering
  	\includegraphics[scale=0.85]{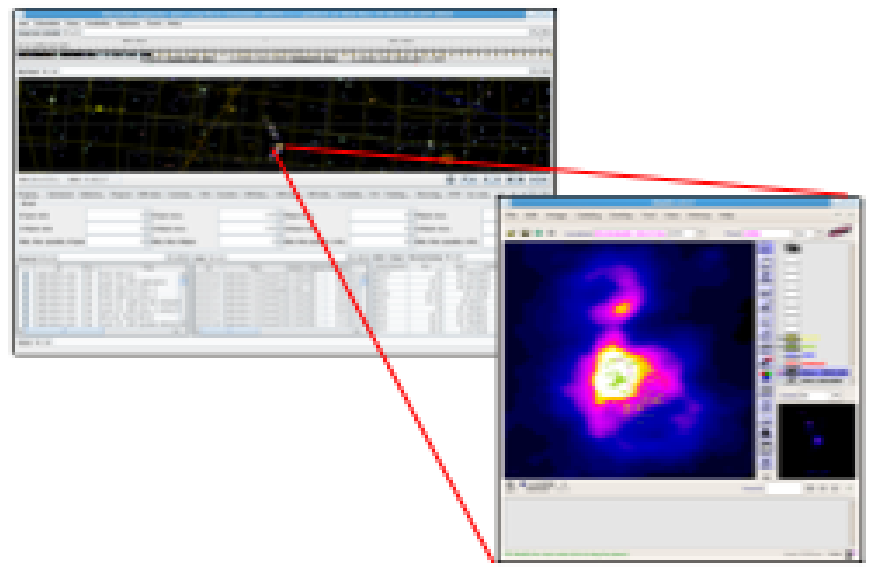}
  	\caption{Joint HILTS-Aladin session centered at M42}
	\label{fig:HiltsAladin}
\end{figure}

\begin{figure}[!ht]
  	\centering
  	\includegraphics[scale=0.40]{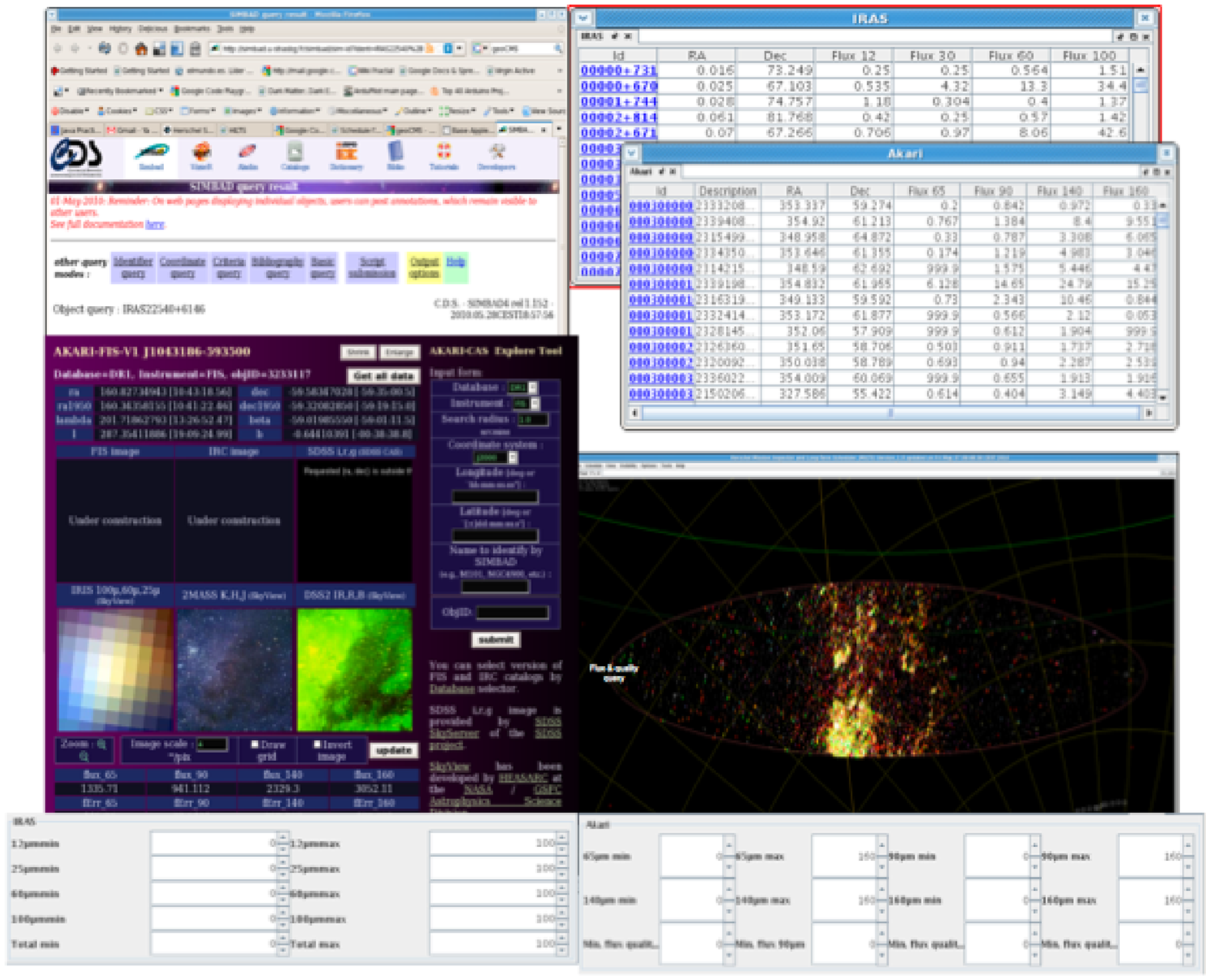}
  	\caption{Catalog capabilities: IRAS and AKARI sources visible at the MGA Herschel constraint are displayed.}
	\label{fig:Catalogs}
\end{figure}

HILTS can also inter-operate with VO tools such as Aladin \cite{F2_adassxx} using the SAMP protocol (see figure \ref{fig:HiltsAladin}): If an HILTS and Aladin session are running simultaneously the former can transmit its current position to the latter where we can analyze in deeper detail the region of interest.

\section{The Scientific Mission Planning System (SMPS)}

The purpose of the SMPS is to generate a daily schedule of telecommands that is uploaded to the spacecraft.  This includes:

\begin{itemize}
	\item commands to the Attitude Control and Monitoring System (ACMS) to maneuver the spacecraft, both to slew between observations and to perform raster and scan patterns
	\item commands to the scientific instruments to perform measurements
	\item commands to perform various engineering operations such as recycling the cryogenic cooler.
\end{itemize}

For scheduling purposes, the mission is divided into operational days (ODs) that cover the period from the start of one DTCP to the start of the next. 
The duration of an operational day is nominally 24 hours, but may vary depending on the active ground station.

\begin{figure}[!ht]
  	\centering
  	\includegraphics[scale=0.32]{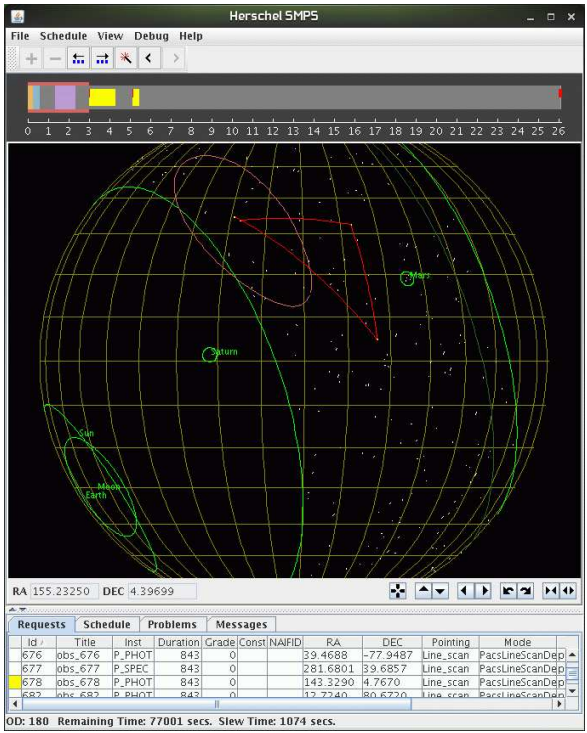}
  	\caption{SMPS main screen}
	\label{fig:mps_main}
\end{figure}

SMPS is a Java tool, whose main screen is divided into a set of views (see figure \ref{fig:mps_main}):

\begin{itemize}
	\item \emph{Time view:} displays the scheduled observations on a time-line, in addition to any temporal constraints on their placement.
	\item \emph{Sky view:} displays the candidate observations that could be scheduled on that OD, plus any spatial constraints on their placement. 
	In addition, it highlights the scheduled observations and draws the slew path that links them.
	\item \emph{Requests table view:} displays the candidate observations as a multi-column table, giving details about each observation.
        \item \emph{Schedule table view:} displays the scheduled observations as a time-ordered sequence.
        \item \emph{Problems panel:} displays any validation problem detected when the user performs a validation or a simulation of the schedule.
	\item \emph{Status panel:} displays general status information.
\end{itemize}

\subsection{Short term mission planning}

The short term mission planning process starts with the processing of an orbit file delivered by the Flight Dynamics team (FD) located at the Mission Operations Centre (MOC) in ESOC (European Space Operations Centre). An Orbit File is delivered to the HSC
once a week, after trajectory optimizations, covering the remaining period to the end of the mission. Use of the ground station is shared with the Planck mission, which may in
turn affect the scheduling of the Herschel DTCP.

The process continues with the loading of a Planning Skeleton File (PSF) delivered by the MOC (one PSF per OD). The PSF is effectively an empty schedule 
that defines time constraints on various operations, such as commanding the instruments and maneuvering the spacecraft. It also includes certain 
key events, such as Acquisition and Loss of signal (AOS/LOS) and the start/end of DTCP. Spacecraft Operations (SOPS) windows are included to reserve time for MOC to insert commands 
for orbit corrections, reaction wheel biasing, etc, when they receive the completed schedule from the HSC.

Scheduling observation of moving Solar System Objects (SSO), such as planets, comets and asteroids, makes use of ephemeris files that are 
obtained from the JPL Horizons system. The SMPS performs various corrections such as proper motion, stellar aberration and boresight alignment and
in the case of SSO, light travel time.

The SMPS is used to add scientific observations and other activities to the schedule. When the Mission Planner is satisfied with the schedule, 
he generates a Planned Observation Sequence (POS) file that expands the observations down to the level of individual telecommands.

\begin{figure}[!ht]
  	\centering
  	\includegraphics[scale=0.35]{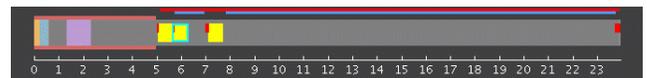}
  	\caption{Operational Day time-line with 3 observations added. The observations are shown in yellow. The slews just before each observation and at the end of the OD are shown in red}
	\label{fig:mps_timeline}
\end{figure}

The POS contains pointing commands that invoke basic ACMS pointing patterns, such as fine pointing, raster and line scan. More complex pointing 
modes are constructed using these basic modes as building blocks. While the spacecraft is executing a pointing pattern, such as a raster, 
the instrument is sent a sequence of telecommands, which must be carefully synchronized with the sequence of spacecraft maneuvers. This 
synchronization is achieved by defining the spacecraft and instrument commands for each observing mode in a special language which models the execution
and timing of operations on the spacecraft.

The Mission Planner provides a summary of the schedule and gives it to the Project Scientist for approval. When it has been approved, 
the mission planner exports the schedule to MOC for further processing and uplink. 

\begin{figure}[!ht]
  	\centering
  	\includegraphics[scale=0.35]{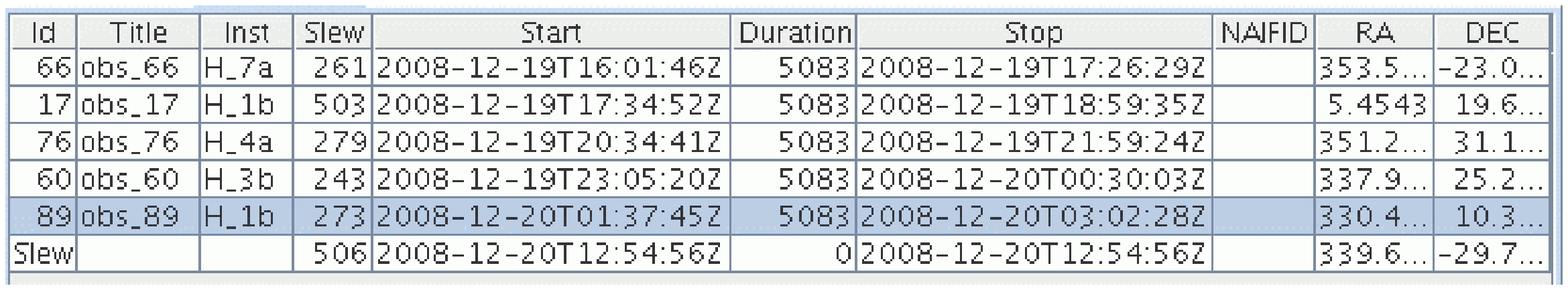}
  	\caption{Example of an schedule with 4 HIFI observations and a final slew}
	\label{fig:mps_schedule}
\end{figure}

When MOC receives the schedule, they insert various commands such as orbit correction maneuvers, into the SOPS windows reserved for them, 
and perform checks on spacecraft constraints and slew times. As a result of this expansion and checking, the POS becomes an Enhanced Planned 
Observation Sequence (EPOS) file. These form the the basis for uplink. If MOC encounter a 
problem, they may notify the HSC and request that the affected days are re-planned.

Each time the SMPS generates a POS file, it commits the changes in the database and marks the observations as 'scheduled' so that they are no 
longer available for scheduling in subsequent ODs. The SMPS does not allow further changes to the schedule unless it is first de-committed. 
This allows the OD to be re-planned and releases the observations so that they can be rescheduled. De-committing a schedule requires a procedural 
interaction with MOC to ensure that schedules are not changed once they have been uplinked to the spacecraft or executed. It also requires 
authorization from the Project Scientist. 

The mission planning process is usually carried out on a weekly basis. MOC deliver a set of seven PSFs at least 15 working days prior to uplink. 
The HSC delivers the corresponding POS files back to MOC at least 10 days before uplink. This allows time for up to two iterations if problems 
are encountered. A shorter turn-around is possible in special cases, such as dealing with a Target of Opportunity (ToO). A ToO may require 
decommitting the affected schedule(s) and replanning.

\subsection{Pointings}

SMPS can display a schematic view of the observation pointing, which does not necessarily include all the complexities of the real observation, 
such as interrupted slews in line scans. The observation does not need to be scheduled, so the orientation is an approximate one for the OD. 
Once an observation is scheduled at a particular time, it is possible to generate a more accurate view using the ACMS simulator.  In this case, 
the attitude evolution for the OD in a sky view is shown. It is possible to zoom in to inspect the details of an individual observation and it 
models subtle details of the ACMS behavior.

\begin{figure}[!ht]
  	\centering
  	\includegraphics[scale=0.35]{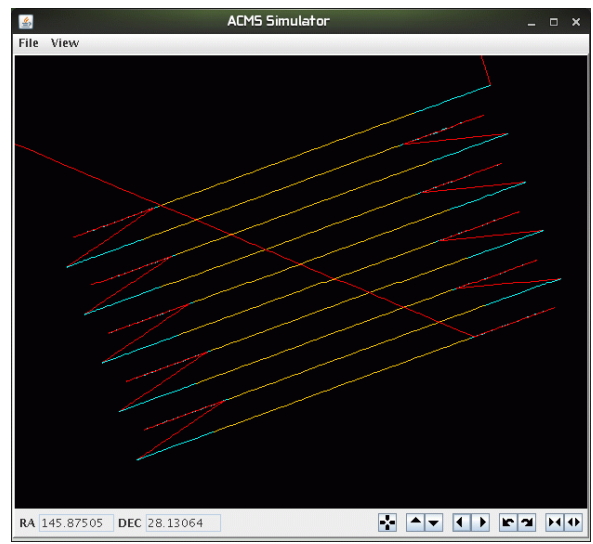}
  	\caption{Detailed simulation of a line-scan}
	\label{fig:mps_simulator}
\end{figure}

Figure \ref{fig:mps_simulator} shows the simulation window zoomed-in to display the details of a line-scan observation. It can be seen that the 
scan slews back to the end of the scan line after deceleration because holds are inserted between the scan lines. The ACMS simulator also performs
a final check of spacecraft commands and constraints.

\section{Conclusions}
The mission planning software used at Herschel Science Centre (HSC) has been presented. It covers the whole range of needed functionality for dealing with the difficult task of 
mission planning and scheduling. HILTS deals with overall mission browsing, inspection, medium/long mission planning and statistical
capabilities. The Scientific mission planning system (SMPS) with the detailed short-term scheduling producing the telecommand 
set to be uplinked to the Herschel satellite.
Both tools have been developed using a common object-oriented framework. This framework has proven to be highly modular and
 potentially reusable for future applications.

\section{Acknowledgments}
We would like to thank the Herschel mission planning team: Alvaro Llorente, Mar Sierra and Veronica Orozco for their constant feedback; 
Leo Metcalfe and Johannes Riedinger for their continuous support to the development.
We would also like to thank I\~{n}aki Ortiz from the Scientific Archives Team for providing some visualization code and useful comments.
Last but not least, we acknowledge the thorough analysis from the two anonymous referees.

\bibliography{author.bib}
\bibliographystyle{aaai}
\end{document}